# Intrinsic high electrical conductivity of stoichiometric SrNbO3 epitaxial thin films

Daichi Oka,[1,2] Yasushi Hirose,[1,3,4,*] Shoichiro Nakao,[3,4] Tomoteru Fukumura,[1,2,4] and Tetsuya Hasegawa[1,3,4]

[1]*Department of Chemistry, School of Science, The University of Tokyo, 7-3-1 Hongo, Bunkyo, Tokyo 113-0033, Japan*
[2]*Department of Chemistry, Graduate School of Science, Tohoku University, 6-3 Azaaoba, Aramaki, Aoba, Sendai 980-8578, Japan*
[3]*Kanagawa Academy of Science and Technology (KAST), 3-2-1 Sakado, Takatsu, Kawasaki 213-0012, Japan*
[4]*CREST, Japan Science and Technology Agency, 7-3-1 Hongo, Bunkyo, Tokyo 113-0033, Japan*

SrVO$_3$ and SrNbO$_3$ are perovskite-type transition-metal oxides with the same $d^1$ electronic configuration. Although SrNbO$_3$ ($4d^1$) has a larger $d$ orbital than SrVO$_3$ ($3d^1$), the reported electrical resistivity of SrNbO$_3$ is much higher than that of SrVO$_3$, probably owing to nonstoichiometry. In this paper, we grew epitaxial, high-conductivity stoichiometric SrNbO$_3$ using pulsed laser deposition. The growth temperature strongly affected the Sr/Nb ratio and the oxygen content of the films, and we obtained stoichiometric SrNbO$_3$ at a very narrow temperature window around 630 °C. The stoichiometric SrNbO$_3$ epitaxial thin films grew coherently on KTaO$_3$ (001) substrates with high crystallinity. The room-temperature resistivity of the stoichiometric film was $2.82 \times 10^{-5}$ cm, one order of magnitude lower than the lowest reported value of SrNbO$_3$ and comparable with that of SrVO$_3$. We observed a $T$-square dependence of resistivity below $T^* = 180$ K and non-Drude behavior in near-infrared absorption spectroscopy, attributable to the Fermi-liquid nature caused by electron correlation. Analysis of the $T$-square coefficient $A$ of resistivity experimentally revealed that the $4d$ orbital of Nb that is larger than the $3d$ ones certainly contributes to the high electrical conduction of SrNbO$_3$.

## I. INTRODUCTION

Perovskite oxides ($ABO_3$) have been extensively studied as a platform for systematic research on the electronic states and physical properties of transition-metal oxides (TMOs). A variety of studies have greatly contributed to the theoretical framework for the electronic states of $3d$ TMOs [1–5]. In contrast, the electronic properties of $4d$ TMOs are less understood, except for a few examples, such as ruthenates [6,7]. It is widely accepted that spatially spread $d$ orbitals in $4d$ TMOs would cause higher electrical conductivity owing to their stronger hybridization with oxygen and a weaker on-site Coulomb interaction than in $3d$ TMOs with an equivalent electronic configuration. This electronic picture has been supported by systematic studies on the optical conductivity and x-ray absorption of SrMO$_3$ ($M = 3d$ or $4d$ transition metal) [8,9].

Among the $4d$ perovskite oxides, however, the intrinsic electrical properties of SrNbO$_3$ ($4d^1$ configuration) have not been revealed. For example, the reported resistivity of SrNbO$_3$ [10–12] is one order of magnitude higher than that of SrVO$_3$ [13,14], a $3d$ TMO with the same $d^1$ configuration as SrNbO$_3$, which runs contrary to the general expectation that $4d$ TMOs should have a higher conductivity than the $3d$ ones. So far, this phenomenon has yet to be explained.

Investigating the electrical properties of SrNbO$_3$ is difficult because it is hard to avoid Sr deficiencies in bulk samples [15,16], though Sr-deficient bulk samples have recently attracted attention because they exhibit both metallic electrical conduction and photocatalytic activity for visible light [10]. One possible cause for the formation of Sr deficiencies is the high processing temperature, over 1000 °C, which is required for bulk synthesis. This problem could be solved by low-temperature epitaxial growth of SrNbO$_3$ thin films: For example, Balasubramaniam *et al.* and Tomio *et al.* reported the epitaxial growth of an SrNbO$_3$ thin film with an almost stoichiometric Sr/Nb ratio by pulsed laser deposition (PLD) at relatively low temperatures of 800 °C [17] and 770 °C [11]. However, the SrTiO$_3$ substrates used in these two studies might have become conductive under reductive growth conditions, which brings into question the reliability of the SrNbO$_3$ transport properties, as we will discuss later. Another problem is the large lattice mismatch between SrNbO$_3$ ($a = 4.023$ Å in a pseudocubic approximation) and SrTiO$_3$ ($a = 3.905$ Å; mismatch = −2.93%), which would degrade the film quality.

---

[*] Electronic address [e-mail]: hirose@chem.s.u-tokyo.ac.jp

In the present study, we grew epitaxial SrNbO$_3$ thin films on single-crystal KTaO$_3$ ($a$ = 3.989 Å; mismatch = −0.85%) using PLD. We found that both the oxygen amount and the Sr/Nb ratio of the films strongly depended on the growth temperature. The stoichiometric SrNbO$_3$ epitaxial thin film grown at an optimized temperature exhibited remarkably high conductivity, comparable with SrVO$_3$, with the temperature dependence as a Fermi liquid.

## II. EXPERIMENT

We grew the films on the (001) plane of KTaO$_3$ substrates (MTI Corporation) under the background pressure of the growth chamber (< 1 × 10$^{-8}$ Torr). To deposit the material, a ceramic pellet of Sr$_2$Nb$_2$O$_7$ was ablated by a KrF excimer laser ($\lambda$ = 248 nm), operated at an energy fluence of 0.33–0.44 J cm$^{-2}$ shot$^{-1}$ and a repetition rate of 10 Hz. The film thickness was 15–40 nm. The substrate temperature was varied using infrared lamp heating.

The crystal structures of the thin films were examined by x-ray diffraction (XRD) using a four-axis diffractometer (Bruker AXS D8 Discover). The cation composition of the films was evaluated by x-ray photoemission spectroscopy (XPS) and inductively coupled plasma mass spectrometry (ICP-MS). Before each XPS measurement, the sample surface was cleaned by Ar ion sputtering. The optical absorption coefficient was obtained from transmittance and reflectance spectra, measured with an ultraviolet-visible near-infrared (NIR) spectrometer. The electrical resistivity, carrier density, and mobility were evaluated by the four-probe method and Hall measurements using a conventional Hall-bar geometry with 100-nm-thick Ag electrodes. Low-temperature measurements were conducted with a physical property measurement system (Quantum Design Model 6000).

## III. RESULTS AND DISCUSSION

XRD measurements revealed that every film possessed a tetragonal perovskite structure, epitaxially grown on the KTaO$_3$ substrate with a cube-on-cube relation. Though the 2$\theta$-$\theta$ pattern showed no impurity phase, the films grown at temperatures lower than 600 °C exhibited a streaky diffraction pattern between the 001 and 101 peaks of the perovskite SrNbO$_3$ in the reciprocal-space image scanned by a two dimensional area detector [Fig. 1(a)]. The 2$\theta$ angle of this streak pattern is around 26°, which corresponds well with the strongest 080 reflection of Sr$_2$Nb$_2$O$_7$ ($a$ = 3.96 Å, $b$ = 5.71 Å, $c$ = 26.76 Å) [18] while it could not be assigned to other possible compounds such as the Ruddlesden-Popper phase. The $\chi$ angle of the streak suggests a (014) orientation of the impurity $Sr_2Nb_2O_7$, which is the same epitaxial orientation of $La_2Ti_2O_7$ grown on the (100) plane of $SrTiO_3$ [19]. Because we used an oxygen-rich target ($Sr_2Nb_2O_7$), excess oxygen layers could be readily formed without introducing oxygen gas. In fact, the streak pattern disappeared at higher growth temperatures, which were more reductive growth conditions [Fig. 1(a)].

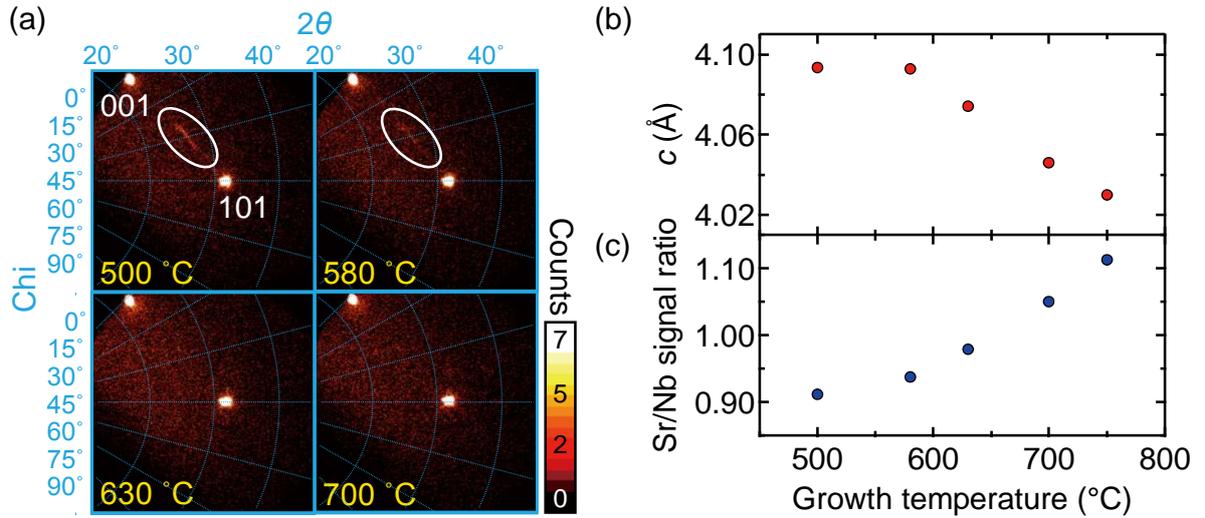

FIG. 1. (a) Two-dimensional detector images of the SrNbO$_3$ thin films grown on KTaO$_3$ single crystals at various temperatures. The tilt angle of the sample ($\chi$) was 45°. The 001 and 101 peaks of the films and substrates overlap with each other. White circles indicate the streaky diffraction patterns from the films. The scale of the color bar representing the intensity is set to make the weak streak pattern clear. The 001 and 101 peaks are displayed with the saturated color, of which counts are over 3000 and 5000, respectively. (b) Out-of-plane lattice constants of the SrNbO$_3$ films plotted against growth temperature. (c) XPS intensity ratios between the Sr 4$p$ and Nb 3$d$ peaks of the SrNbO$_3$ thin films grown at various temperatures. These values are normalized by the photoionization cross section.

While the excess oxygen layer disappeared at higher temperatures, the lattice constant, calculated from 002 and 004 diffractions, decreased as the growth temperature increased above 600 °C [Fig. 1(b)]. This reduction was probably caused by cation off-stoichiometry: An XPS measurement revealed that the Sr/Nb ratio monotonically increased with increasing growth temperature [Fig. 1(c)]. An ICP-MS measurement confirmed that the Sr/Nb ratio of the film grown at 630 °C was 0.96 + 0.10, an essentially stoichiometric value within the experimental error. Because XRD confirmed the phase-pure perovskite structure, even at high growth temperatures, the Sr-rich compounds were likely created by the introduction of an Nb vacancy at the $B$ site or the partial substitution of Sr in the $B$ site [20]. From these results, we conclude that the optimal temperature to grow $SrNbO_3$ epitaxial thin films is 630 °C. Hereafter we refer to the film grown at the optimal temperature as the stoichiometric $SrNbO_3$ epitaxial thin film.

We investigated the crystal structure of the stoichiometric $SrNbO_3$ epitaxial film in more detail using high-resolution XRD. Figure 2(a) shows a high-resolution $2\theta$-$\theta$ pattern of the stoichiometric $SrNbO_3$ thin film. The fringe pattern evolved around the 002 peak indicates that the film surface and film/substrate interface were both flat [inset of Fig. 2(a)]. Figure 2(b) shows a reciprocal-space map around the 103 peaks of the film and the substrate. The film was coherently grown on the substrate, and the out-of-plane lattice constant was evaluated as 4.07 Å, which indicates compressive epitaxial strain from the substrate. A rocking curve of the 002 peak had a full width at half maximum as small as 0.04°, indicating that the film had good crystallinity.

Figure 2(c) shows an absorption coefficient spectrum of the stoichiometric $SrNbO_3$ epitaxial thin film, revealing broad absorption in both the visible and NIR regions. A Tauc plot [inset of Fig. 2(c)] was made, assuming an indirect band-to-band transition, based on a previous band calculation [10]. The plot shows a dip structure around 2 eV, suggesting an overlap of absorption bands with different origins in the visible and NIR regions. The absorption in the visible region corresponds well to that in Sr-deficient bulk $SrNbO_3$, which has been attributed to a transition from the conduction band to a higher-level unoccupied band [10]. On the other hand, the NIR absorption is certainly derived from the high density of electrons, discussed later. The shape of the spectrum in the NIR region was not reproduced well with a simple Drude model [dashed line in Fig. 2(c)], suggesting considerable electron correlation. An extended Drude model may be necessary to explain the spectrum, similar to the case in other electron correlated systems such as $Sr_2RuO_4$ [21].

Figure 3 compares the resistivity, carrier (electron) density, and mobility measured at 300 K from $SrNbO_3$ thin films grown at various temperatures. The stoichiometric $SrNbO_3$ epitaxial thin film exhibited the minimum resistivity [Fig. 3(a)], and its carrier density agreed with the nominal value assuming that each Nb ion supplies one electron [dashed line in Fig. 3(b)]. Notably, the minimum resistivity of the stoichiometric $SrNbO_3$ is $2.82 \times 10^{-5}$ $\Omega$cm, which is one order of magnitude lower than previously reported values [10,11]. As the growth temperature deviated from the optimal value, the resistivity drastically increased, probably owing to the off-stoichiometry mentioned previously: Excess oxygen layers introduced at lower growth temperatures compensate the electron carriers and suppress their mobility by breaking up the $NbO_6$ octahedral network. At higher growth temperatures, though, other defects are generated, such as Nb vacancies and Sr substitutions in the *B* site [20], both of which increase the Nb valence from 4+ to 5+ and thus decrease the carrier density.

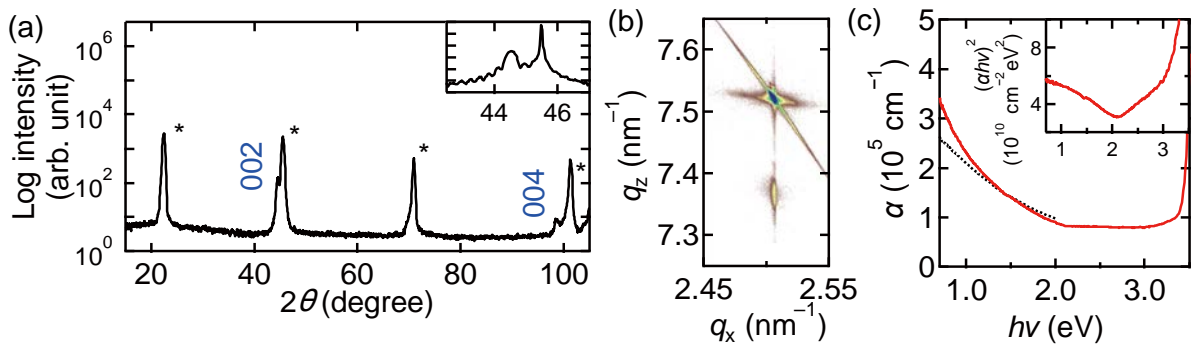

FIG. 2. (a)XRD$2\theta$-$\theta$ pattern of the stoichiometric $SrNbO_3$ epitaxial thin film. The inset shows the pattern near the 002 peak taken with a high-resolution detector. (b) XRD reciprocal space map image around the 103 diffractions of the $SrNbO_3$ thin film and $KTaO_3$ substrate. (c) Absorption coefficient spectrum of the stoichiometric $SrNbO_3$ epitaxial thin film. The dashed line is a fit to the simple Drude model, using the carrier density evaluated by a Hall measurement. The inset shows a Tauc plot that assumes an indirect band gap.

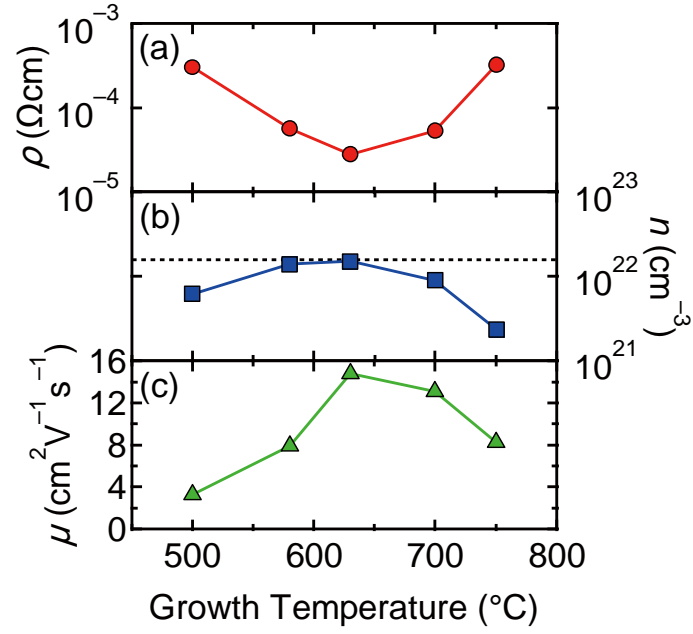

FIG. 3. Growth temperature dependence of (a) resistivity $\rho$, (b) carrier density $n$, and (c) Hall mobility $\mu$ of the SrNbO$_3$ thin films measured at room temperature. The dashed line in (b) indicates the nominal carrier concentration.

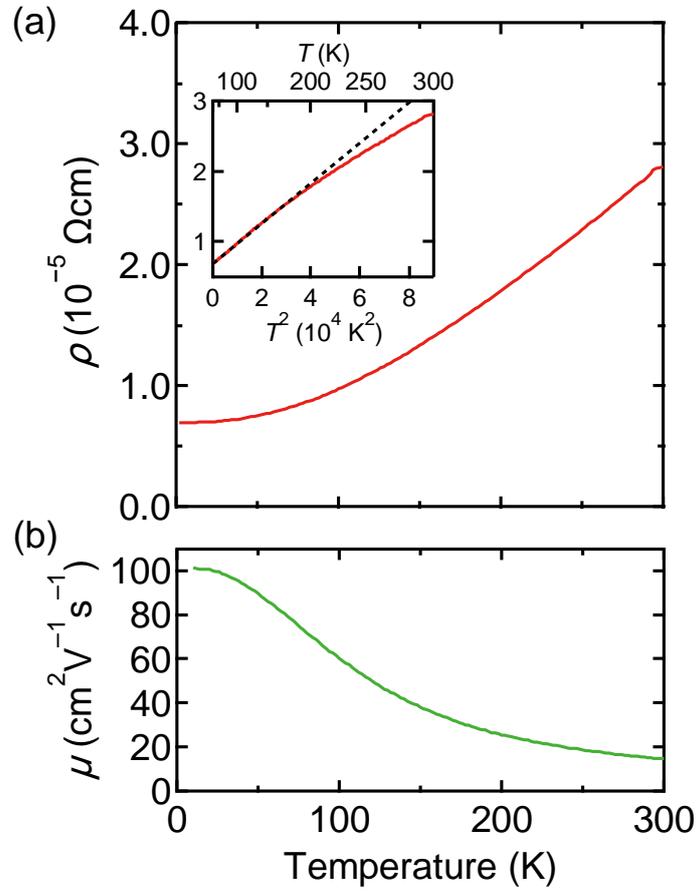

FIG. 4. Temperature dependence of the (a) resistivity and (b) Hall mobility of the stoichiometric SrNbO$_3$ epitaxial thin film. The inset in (a) shows the resistivity plotted against $T^2$. The dashed line is a linear fit of the curve in the low-temperature region.

Figure 4 plots the resistivity and mobility of the stoichiometric $SrNbO_3$ epitaxial thin film as functions of temperature. The resistivity exhibited a metallic temperature dependence ($d\rho/dT > 0$) from room temperature to 2 K. The mobility tended to increase with decreasing temperature and reached over 100 cm$^2$ V$^{-1}$ s$^{-1}$ at around 20 K. Tomio *et al.* investigated the transport properties of an $SrNbO_3$ thin film grown on an $SrTiO_3$ substrate and found resistivities of $1 \times 10^{-6}$ Ωcm at 40 K and $2.7 \times 10^{-4}$ Ωcm at 300 K [11]. The former resistivity is about one order of magnitude lower than that of our film, while the latter is about one order of magnitude higher than ours (Fig. 4). One possible explanation for this discrepancy is that the $SrTiO_3$ substrate used by Tomio *et al.* contributed greatly to the electrical conduction, particularly at low temperatures, owing to Nb interdiffusion at the film/substrate interface, the formation of oxygen vacancies, or both. In contrast, the $KTaO_3$ substrate in the present study does not show conductivity by Nb doping, and it requires much more reductive conditions to introduce oxygen vacancies. Thus, we believe that the transport data in Fig. 4 represent the intrinsic properties of $SrNbO_3$.

As shown in the inset of Fig. 4(a), the resistivity of $SrNbO_3$ at low temperatures follows the temperature dependence of $\rho = \rho_0 + AT^2$. Such a *T*-square dependence of $\rho$ is observed in some metallic oxides and is usually attributed to the Fermi-liquid behavior of quasiparticles caused by electron correlation [14,22–28]. In order to verify this attribution in $SrNbO_3$, we compare the *T*-square coefficient *A*, which encapsulates the contribution of electron correlation to resistivity, with those of other Fermi-liquid perovskite oxides. Though Fermi-liquid materials have been widely investigated based on the Kadowaki-Woods rule, $A/\gamma^2$ = const [29,30], where *γ* is the *T* linear coefficient of specific heat, the evaluation of *γ* is difficult for a thin film. Therefore, we propose another empirical relationship between *A* and conducting carrier density *n*. According to Ref. [31], *A* is expressed as follows,

$$A = \frac{4\pi^2 k_B^2}{e^2 \hbar^2} \frac{m_b}{n} \Phi, \quad (1)$$

where $k_B$ is the Boltzmann constant, *e* is elementary charge, *ℏ* is the Planck constant divided by $2\pi$, $m_b$ is the band mass, which would be calculated using the localized density approximation, and Φ is the coefficient of the scattering rate $\Gamma(T,\omega) = \Gamma_0 + \Phi[(2\pi T)^2 + \omega^2]$. Thus, the factor $m_b\Phi$ reflects band dispersion and many body effects in a correlated system. In Eq. (1), $A$ is inversely proportional to $n$, assuming that $m_b\Phi$ is independent of $n$. Indeed, the ln $A$ vs ln $n$ plot for various Fermi-liquid perovskite oxides such as $SrTi_{1-x}Nb_xO_3$, $Sr_{1-x}La_xTiO_3$, $Ca_{1-x}Y_xTiO_3$, and $K_{1-x}Ba_xTaO_3$ shows the linear relationship with a slope of −1 [Fig. 5(a)]. Notably, all the plots are located on a single narrow band with a slope of −1, including other Fermi-liquid perovskite oxides such as $Ca_{1-x}Sr_xVO_3$ and $SrMoO_3$ [shaded area in Fig. 5(a)]. This fact empirically indicates that $m_b\Phi$, i.e., prefactor of the $1/n$, falls in a relatively narrow range in the case of Fermi-liquid perovskite oxides. Because $A$ of the $SrNbO_3$ epitaxial thin film agrees well with this relationship, we considered that $SrNbO_3$ is a Fermi-liquid material: In other words, electron correlation plays an important role in its electrical conduction. The non-Drude behavior observed in the NIR absorption of $SrNbO_3$ also supports this attribution.

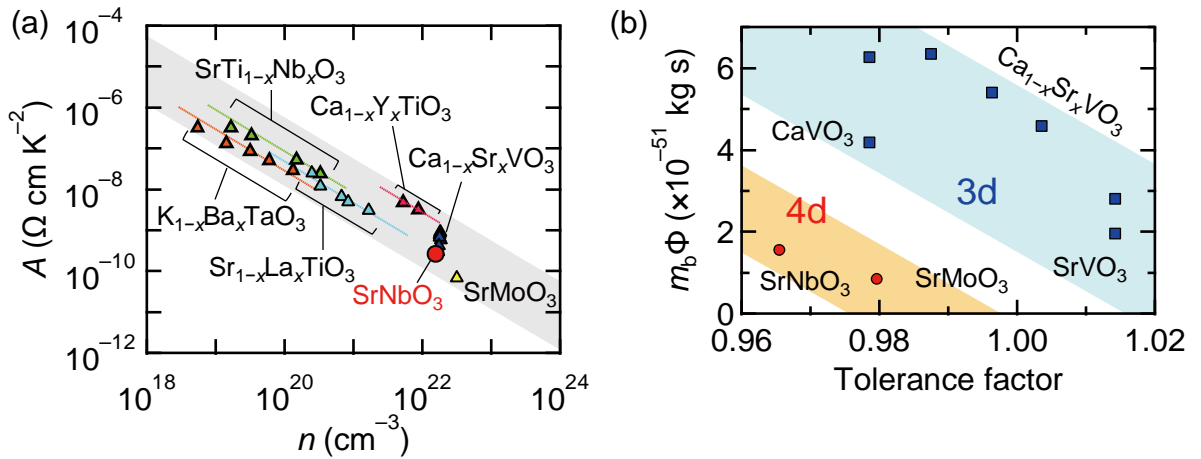

FIG. 5. (a) Plot of the *T*-square coefficient *A* of resistivity vs carrier density *n* for single crystals or epitaxial films of perovskite-type transition-metal oxides [$Sr_{1-x}La_xTiO_3$ [22], $SrTi_{1-x}Nb_xO_3$ [23], $K_{1-x}Ba_xTaO_3$ [24], $Ca_{1-x}Y_xTiO_3$ [25], $CaVO_3$ [26], $SrVO_3$ [14], $Ca_{1-x}Sr_xVO_3$ [27], $SrMoO_3$ [28], and $SrNbO_3$ (this study)]. Dashed lines with a slope of −1 are drawn for several series of materials as guides for the eyes. The gray shaded area indicates the band with a slope of −1 from $m_b\Phi = 5 \times 10^{-52}$ to $2 \times 10^{-50}$ kg s. (b) Plots of the $m_b\Phi$ value vs the tolerance factor for $AMO_3$ (*A* = Sr, Ca, *M* = V, Nb, Mo). Data are taken from the same references as (a). The tolerance factor was calculated from the ionic radius reported in Ref. [32].

For further discussion on the electron correlation in $3d$ and $4d$ transition-metal oxides, we compared the $m_b\Phi$ values of $AMO_3$ ($A$ = Ca, Sr, $M$ = V, Nb, Mo) in more detail. It is known that both band mass and electron correlation in perovskite oxides are affected by the structural distortion and size of the $d$ orbital through the transfer integral. In order to extract the orbital-size contribution to $m_b\Phi$, we plotted $m_b\Phi$ values of $AMO_3$ against the Goldschmidt's tolerance factor $t$ [Fig. 5(b)]. It is known that smaller $t$ increases the electron correlation in perovskite oxides through a decrease in the orbital overlap. This trend is confirmed by the plots of $Ca_{1-x}Sr_xVO_3$, where $m_b\Phi$ increases by decreasing $t$. On the other hand, $m_b\Phi$ values of $SrNbO_3$ and $SrMoO_3$ are clearly smaller than V-based perovskite oxides with similar $t$, indicating that the larger orbital of $4d$ transition metals lead to a smaller electron correlation and a smaller $m_b\Phi$ value. As a result of the balance between the structural distortion and orbital size, $SrMO_3$ ($M$ = V, Nb, Mo) possesses a close $m_b\Phi$.

The $T$-square coefficient $A$ also gives insight into the room-temperature resistivity of $SrMO_3$. The characteristic temperature $T^*$ of $SrNbO_3$, the temperature at which its resistivity began to deviate from the parabolic relation described previously, was ~180 K [Fig. 4(a)], which is comparably high to that of $SrMoO_3$ [28]. Since such high $T^*$ values make the second term in the equation $\rho = \rho_0 + AT^2$ dominant at high temperatures, the close $A$ values of $SrMO_3$ ($M$ = V, Nb, Mo) lead to a comparable resistivity of these materials at 300 K as long as the sample quality is good enough (Table I).

TABLE I. Resistivities at 300K among $SrMO_3$ ($M$ = V, Nb, Mo) as a bulk polycrystal, bulk single crystal, and epitaxial thin film.

| Sample | $\rho$ at 300 K (μΩ cm) | | |
| --- | --- | --- | --- |
| | Polycrystal | Single crystal | Epitaxial thin film |
| $SrNbO_3$ ($4d^1$) | 500 [10] | | 28 (present study) |
| $SrMoO_3$ ($4d^2$) | | 5.1 [25] | 29 [33] |
| $SrVO_3$ ($3d^1$) | 120 [34] | 26 [13] | 25 [14] |

## IV. CONCLUSION

We fabricated $SrNbO_3$ epitaxial thin films on a $KTaO_3$ substrate by PLD. We found that both the cation and anion compositions were very sensitive to growth temperature and that almost stoichiometric $SrNbO_3$ could be obtained at an optimum growth temperature of 630 °C. The stoichiometric $SrNbO_3$ film showed a room-temperature resistivity of $2.82 \times 10^{-5}$ Ωcm, which is one order of magnitude lower than previously reported values and comparable with those of $3d^1$ $SrVO_3$ and $4d^2$ $SrMoO_3$. This result did not agree with a naive expectation that $SrNbO_3$ would show lower resistivity than $SrVO_3$ and $SrMoO_3$ because $Nb^{4+}$ has the largest $d$ orbital among these three materials. Based on the Fermi-liquid behavior of the low-temperature resistivity and the non-Drude feature of the NIR absorption, we inferred that electron correlation plays a significant role in the electrical conduction of $SrNbO_3$. A comparison of the $T$-square coefficient $A$ of $3d$ and $4d$ perovskite oxides revealed an orbital-size contribution to the electron correlation in these materials.

## ACKNOWLEDGMENTS

This study was supported by the Ministry of Education, Culture, Sports, Science and Technology of Japan (MEXT) as part of KAKENHI No. 24760005 and by the Japan Society for the Promotion of Science (JSPS) KAKENHI (Grant No.

248258). A part of this work was conducted in the Research Hub for Advanced Nano Characterization, The University of Tokyo, under the support of the "Nanotechnology Platform" (Project No. 12024046) by MEXT.